\documentclass[
 reprint,
 amsmath,amssymb,
 aps,
]{revtex4-2}

\usepackage{graphicx}
\usepackage{newtxmath}
\usepackage{bm}

\usepackage[colorlinks=true, linkcolor=blue,anchorcolor=blue, citecolor=blue,urlcolor=blue]{hyperref}

\usepackage[normalem]{ulem}

\usepackage{xcolor}

\begin{document}
\preprint{APS/123-QED}

\title{Ferroelectric-controllable spin-orbit torque in two-dimensional multiferroic heterostructure}
\author{Weiyi Pan$^{1}$}
\email{Weiyi.Pan@physik.uni-regensburg.de}
\author{Gusthavo M. S. Brizolla$^{1}$}
\author{Jaroslav Fabian$^{1,2}$}

\affiliation{$^{1}$Institute for Theoretical Physics, University of Regensburg, 93040 Regensburg, Germany\\
$^{2}$Halle-Berlin-Regensburg Cluster of Excellence CCE, University of Regensburg, 93040 Regensburg, Germany\\
}

\begin{abstract}
Spin–orbit torque (SOT), which enables electrical control of magnetization, plays a crucial role in the development of next-generation spintronic devices. Realizing SOT in two-dimensional van der Waals systems, together with achieving efficient nonvolatile manipulation via ferroelectricity, would be highly beneficial for the implementation of tunable logic devices with enhanced storage density. In this work, based on first-principles calculation and using a multiferroic Fe$_{3}$GeTe$_{2}$/In$_{2}$Se$_{3}$ heterostructure as a representative example, we demonstrate that switching the ferroelectric polarization of the In$_{2}$Se$_{3}$ layer induces a pronounced modification in the magnetization-dependent distribution of torkance within the heterostructure. Specifically, when the magnetization is in the plane, where the torque is maximal, reversing the polarization of In$_{2}$Se$_{3}$ from upward to downward enhances the total torkance to more than 150\% of its original value. This substantial variation primarily originates from the polarization-induced modulation of the $z$ component of the time-reversal-odd torkance, which is mainly associated with an approximately 233\% change in the atomic-resolved torque contributed from the middle Fe layer in Fe$_{3}$GeTe$_{2}$ layer. Further analysis reveals that the electronic states near $\Gamma$ on the Fermi surface undergo significant reconstruction upon polarization switching, which is responsible for the observed variation in the time-reversal-odd torque. Our results not only provide new insights into the functional potential of van der Waals multiferroic heterostructures, but also offer a viable strategy for achieving electrically tunable SOT, paving the way for future programmable spintronic devices.

\end{abstract}
\maketitle
\section{Introduction}
Controlling magnetic states using purely electrical methods is essential for spintronics\cite{RevModPhys.76.323}, especially for the development of energy-efficient and non-volatile spintronic memories, in which magnetization switching is used to encode binary information\cite{RevModPhys.96.015005,grollier2020neuromorphic,9427163}. An effective approach to achieve this goal is current-induced spin–orbit torque (SOT)\cite{RevModPhys.91.035004,9427163,miron2011perpendicular,SOT1,SOT2,SOTs}. Enabled by spin–orbit coupling (SOC) and broken centrosymmetry, SOT facilitates the transfer of angular momentum from the crystal lattice to the magnetization\cite{transfer}, thereby offering a promising route toward scalable magnetic random-access memory and spintronic devices for next-generation computing. 

The vector components of SOT are commonly classified into two distinct contributions. One component is an even function of the magnetization satisfying $\textbf{T}^{\textup{even}}(-\textbf{M})$ = $\textbf{T}^{\textup{even}}(\textbf{M})$, which is usually believed to be associated with spin Hall effect\cite{SHE} and its lowest order term in $\textbf{M}$ is termed as damping-like torque, $\textbf{T}_{\textup{DL}} = t_{\textup{DL}} \textbf{M} \times ( \boldsymbol{\sigma} \times \textbf{M} )$, where $\boldsymbol{\sigma} = \hat{z}  \times \textbf{E
}$ is a vector which relates with the direction of electric field. The other component is an odd function of the magnetization satisfying $\textbf{T}^{\textup{odd}}(-\textbf{M})$ = $-\textbf{T}^{\textup{odd}}(\textbf{M})$, which is usually believed to be associated with the interfacial Rashba–Edelstein effect\cite{EDE,johansson2024theory} and its lowest-order term in $\textbf{M}$ is named as field-like torque, $\textbf{T}_{\textup{FL}} = t_{\textup{FL}} \boldsymbol{\sigma} \times \textbf{M} $. From the viewpoint of magnetization dynamics, the time-reversal-even damping-like torque acts like an effective magnetic damping and drives the magnetization towards the direction of current-induced spin polarization. Meanwhile, the time-reversal-odd field-like torque acts like an effective magnetic field and initiates precession of magnetization around spin polarization\cite{guimaraes2020spin}.

\begin{figure*}[ht]
\includegraphics[scale = 0.32 ]{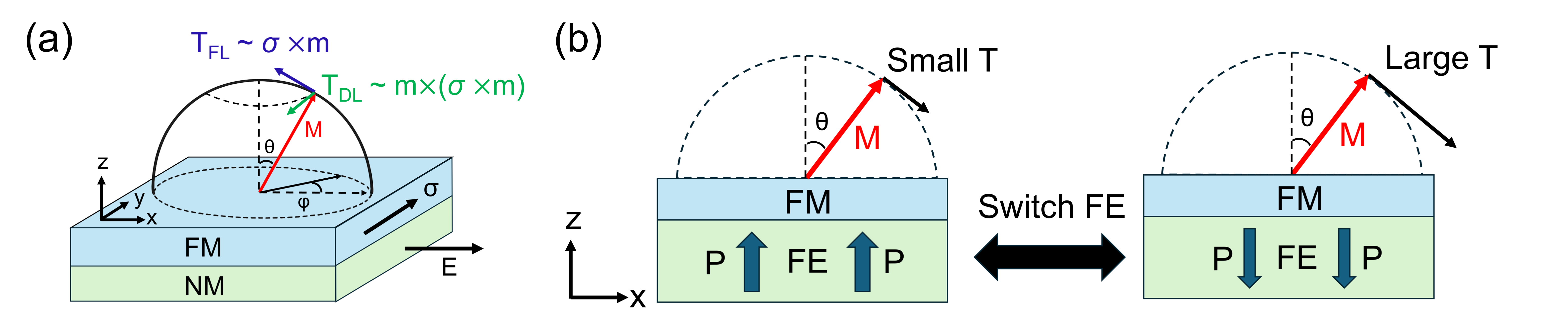}
\caption{\label{1} (a) A Schematic illustration of current-induced spin-orbit torque in FM/NM heterostructure system.  A longitudinal electron current parallel with $\textbf{E}$ in NM layer would induce both damping-like torque $\textbf{T}_{\textup{DL}}$ and field-like torque $\textbf{T}_{\textup{FL}}$. The $\theta$ and $\varphi$ denotes the polar angle and azimuthal angle for the magnetization, respectively. (b) Schematic illustration of ferroelectric-controllable torque in multiferroic heterostructures. By switching the polarization in ferroelectric layer, the magnitude of torque would be altered.}
\end{figure*}

A typical structural platform for realizing SOT is a heterostructure composed of a ferromagnetic (FM) layer and a nonmagnetic (NM) layer, which is shown in Fig. \ref{1}(a). When an electric current flows through such a heterostructure, the strong SOC in the NM layer can generate non-equilibrium spin polarization via the spin Hall effect or the interfacial Rashba–Edelstein effect. This spin polarization exerts a torque on the intrinsic magnetic moments in the FM layer, thereby giving rise to SOT. Experimentally, SOT is commonly realized in metallic multilayer heterostructures consisting of nonmagnetic heavy metals and ferromagnetic metals\cite{HM1,HM2,SOT1,SOT2,HM3,HM4}. However, SOT in metallic multilayer systems is highly sensitive to unavoidable atomic intermixing at the interface\cite{Mixing2,Mixing4}, which may induce unpredictable changes in SOT efficiency and thus pose challenges for reliable device fabrication\cite{Mixing,Mixing3}. Moreover, it is generally challenging to precisely manipulate the atomic and electronic structures of such NM/FM metallic systems in experiments, which further hinders the exploration of novel SOT-related physical phenomena and the development of next-generation functional spintronic devices.

Beyond conventional metallic multilayer systems, emerging two-dimensional (2D) van der Waals (vdW) materials have opened new avenues for realizing highly efficient and tunable SOT\cite{vdW}. Compared with metallic heterostructures, vdW systems can host atomically clean and well-defined interfaces, enabling more stable and reproducible SOT performance and thereby offering significant advantages for device applications. In addition, the intrinsic electronic structures and magnetic properties of 2D all-vdW materials can be readily manipulated through various approaches, such as electrostatic gating\cite{LDA2,huang2018electrical,Gat1,Gat2}, twisting\cite{Twist,Twist2,Twist3,Twist4,Twist5}, and strain engineering\cite{Strain,strain2,strain3}, providing versatile and effective routes for SOT control. To realize SOT in vdW systems, a typical experimental configuration involves placing a monolayer 2D FM with perpendicular magnetic anisotropy—commonly Fe$_{3}$GeTe$_{2}$\cite{LDA2,FGT11}, Fe$_{3}$GaTe$_{2}$\cite{LDA,FGaT,FGaT2} , and CrTe$_{2}$\cite{CrTe,CrTe22}—on a vdW nonmagnetic substrate with strong SOC. In such heterostructures, the combination of strong SOC and interface-induced inversion asymmetry gives rise to the emergence of SOT. Moreover, proximity coupling between the 2D magnetic layer and the adjacent nonmagnetic layer can modulate the charge–spin conversion process, potentially leading to exotic SOT phenomena. Experimentally, energy-efficient SOT operation at room temperature has been demonstrated in 2D FM/topological insulator heterostructures\cite{TI,TI2,TI3}, where topological interfacial states are believed to play a crucial role in enhancing SOT efficiency\cite{TI4}. Furthermore, experimentally accessible heterostructures composed of 2D ferromagnets and vdW substrates with low crystal symmetry, such as WTe$_{2}$\cite{WT1,WT2,WTe4,WTE3} and TaIrTe$_{4}$\cite{TTI1,TTI2,TTI3,TTI4}, have enabled field-free switching of perpendicular magnetization in all-vdW systems due to the existence of out-of-plane damping-like torque \cite{WT1,WT2,WTe5,TTI1,TTI2}, which is essential for achieving high-density spintronic devices. From the view 
point of theory, only few magnetic vdW syetems, such as Janus CrTeSe\cite{Theo1}, Graphene/CrSBr heterostructure\cite{THEO2}, graphene sandwiched by CrGeTe$_{3}$ and WS$_{2}$\cite{theo3}, CrI$_{3}$/TaS$_{2}$ heterostructure\cite{theo4}, as well as monolayer Fe$_{3}$GeTe$_{2}$\cite{FGT}, are recently predicted to host SOT. Despite the above mentioned advances, the realization of SOT in fully van der Waals materials still remains relatively scarce, and the exploration of additional candidate systems and understanding its microscopic origin from the viewpoint of first-principles calculation is therefore highly desirable.

In addition to extending the realization of SOT to a broader class of vdW materials, another crucial objective is to achieve nonvolatile electrical control of SOT\cite{EE1,EE2,EE3,EE4}. Such capability is highly desirable for the development of high-density, electrically programmable logic devices\cite{ESOT,ESOT2,ESOT3} and neuromorphic systems driven by electrical stimuli\cite{grollier2020neuromorphic,mukherjee2023graphene}. One promising platform for realizing nonvolatile SOT manipulation is vdW multiferroic heterostructures\cite{MF,MF2}, which typically consist of a 2D FM layer coupled to a 2D ferroelectric substrate, as shown in Fig. \ref{1}(b). In these vdW multiferroic heterostructures, switching the electric polarization of the ferroelectric layer modifies the interfacial coupling between the ferroelectric and magnetic layers. This interfacial modulation subsequently alters the electronic structure and magnetic properties of the system, giving rise to a variety of ferroelectric-switchable physical phenomena, including electronic topology\cite{MF3,MF4}, magnetic skyrmions\cite{MF7,MF8,MF9,MF16,MF17}, valleytronics\cite{MF5,MF6,MF12}, magnetic anisotropy\cite{MF,MF10,MF11,MF17,MF18}, and tunneling magnetoresistance\cite{MF13,MF14,MF15}. Given that SOT is intrinsically linked to the electronic properties of a material system\cite{Kubo1,PhysRevB.95.134449,PhysRevB.93.224420}, one may naturally expect that polarization switching can induce nonvolatile modulation of SOT in vdW multiferroic heterostructures. From another perspective, reversal of the ferroelectric polarization directly modulates the spin–orbit fields and charge–spin conversion processes in the system\cite{gkt2-x7mm,PhysRevB.107.165140,PhysRevMaterials.6.L091404,pan2026tunableedelsteineffectintrinsic,milivojevic2026ferroelectric,PhysRevB.100.155408}, which in turn leads to variations in the magnitude and symmetry of SOT. Nevertheless, systematic studies on ferroelectrically tunable SOT in realistic vdW multiferroic heterostructures, as well as a clear understanding of the underlying microscopic mechanisms, remain limited. Further theoretical and experimental investigations are therefore highly warranted.

In this work, we report ferroelectrically controllable SOT in a 2D vdW multiferroic heterostructure, taking the Fe$_{3}$GeTe$_{2}$/In$_{2}$Se$_{3}$ heterobilayer as a representative example. Based on first-principles calculations, we demonstrate that reversing the ferroelectric polarization of In$_{2}$Se$_{3}$ induces a substantial modulation of torkance in the heterostructure. Specifically, when the magnetization is oriented along the in-plane direction, switching the polarization from the upward to the downward state leads to the SOT reaching more than 150\% of its original value. This pronounced variation primarily originates from the polarization-induced modulation of the $z$-component of time-reversal-odd torkance. Further analysis reveals that polarization switching significantly alters the electronic structure on the Fermi surface, which provides a microscopic understanding of the polarization-dependent SOT modulation. Our results thus establish an effective strategy for realizing nonvolatile and electrically controllable SOT in realistic 2D vdW systems, thereby paving the way toward tunable and energy-efficient spintronic devices.

\section{Method}
First-principles calculations were performed within the framework of density functional theory (DFT) using the Quantum Espresso package\cite{QE}. A Monkhorst–Pack $k$-point mesh of $18 \times 18 \times 1$ was employed, and the kinetic energy cutoffs for the plane-wave basis set and charge density were set to 90 Ry and 800 Ry, respectively. Projector-augmented-wave (PAW)\cite{PAW} pseudopotentials together with the local density approximation (LDA) exchange–correlation functional were adopted, which have been demonstrated to accurately describe the magnetic properties of Fe$_{3}$GeTe$_{2}$\cite{LDA,LDA2,LDA3}. Structural relaxations were carried out until the residual forces on each atom were smaller than 3 × 10$^{-4}$ Ry/$a_{0}$, while the total energy convergence criterion for the self-consistent calculations was set to 10$^{-8}$ Ry, where $a_{0}$ denotes the Bohr radius. To account for the interlayer coupling between the Fe$_{3}$GeTe$_{2}$ and In$_{2}$Se$_{3}$ monolayers, the semi-empirical Grimme DFT-D2 van der Waals correction was employed\cite{DFT-D3}. A vacuum layer of approximately 20 \AA was introduced along the $z$ direction to eliminate spurious interactions between periodic images.

After completing the DFT calculations, symmetry-adapted Wannier functions were constructed for the heterostructure systems using the Wannier90\cite{w901} and WannierBerri packages\cite{SAWF,tsirkin2021high}. During the wannierization procedure, the following atomic orbitals were chosen as projection bases: $d$ orbitals for Fe atoms, $p$ orbitals for Ge, Te, and Se atoms, and both $s$ and $p$ orbitals for In atoms. Based on the resulting Wannier Hamiltonian, the torkance tensor $\boldsymbol{\tau}$ can be evaluated, which denotes the torque per electric field strength and thus relates the torque $\textbf{T}$ to the magnetization direction $\textbf{M}$ and applied electric field $\textbf{E}$ through\cite{Kubo1}
 \begin{equation}
     \textbf{T}(\textbf{M}) = \boldsymbol{\tau}(\textbf{M}) \textbf{E}.
 \end{equation}

 Within the framework of linear-response theory, the torkance tensor can be generally decomposed into a time-reversal-even component, corresponding to the interband contribution to the torkance, which is an even function of magnetization and is given by\cite{Kubo1,FGT}:

\begin{equation}
\begin{aligned}
\tau_{ij}^{\mathrm{even}}
={}&
\frac{e\hbar}{2\pi N}
\sum_{\mathbf{k},\,n\neq m}
\operatorname{Im}
\left[
\langle n\mathbf{k}|\mathcal{T}_i|m\mathbf{k}\rangle
\langle m\mathbf{k}|v_j|n\mathbf{k}\rangle
\right]
\\
&\times
\Bigg\{
\frac{
\Gamma\left(E_{\textup{m\textbf{k}}}-E_{\textup{n\textbf{k}}}\right)
}{
\left[
\left(E_{\textup{F}}-E_{\textup{n\textbf{k}}}\right)^2+\Gamma^2
\right]
\left[
\left(E_{\textup{F}}-E_{\textup{m\textbf{k}}}\right)^2+\Gamma^2
\right]
}
\\
&\qquad
+
\frac{
2\Gamma
}{
\left(E_{\textup{n\textbf{k}}}-E_{\textup{m\textbf{k}}}\right)
\left[
\left(E_{\textup{F}}-E_{\textup{m\textbf{k}}}\right)^2+\Gamma^2
\right]
}
\\
&\qquad
+
\frac{
2
}{
\left(E_{\textup{n\textbf{k}}}-E_{\textup{m\textbf{k}}}\right)^2
}
\operatorname{Im}
\ln
\frac{
E_{\textup{m\textbf{k}}}-E_{\textup{F}}-i\Gamma
}{
E_{\textup{n\textbf{k}}}-E_{\textup{F}}-i\Gamma
}
\Bigg\}.
\end{aligned}
\end{equation}

\begin{figure*}[ht]
\includegraphics[scale = 0.35 ]{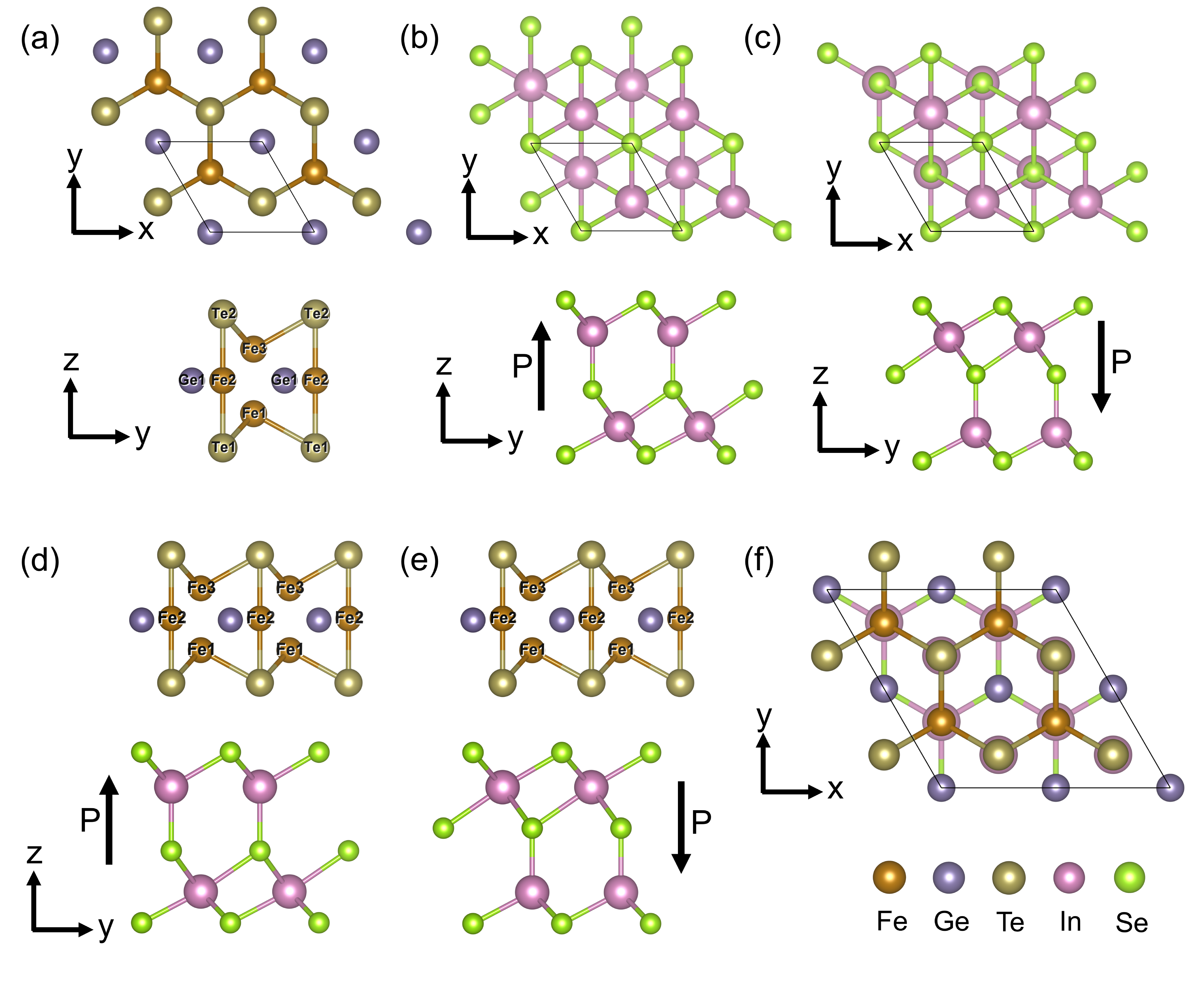}
\caption{\label{2} (a) Atomic structure of Fe$_{3}$GeTe$_{2}$ monolayer. Distinct atoms are labeled here. (b) and (c) denotes the atomic structure of monolayer In$_{2}$Se$_{3}$ with polarization pointing upwards and downwards, respectively. (d) and (e) denotes the sideview of Fe$_{3}$GeTe$_{2}$/In$_{2}$Se$_{3}$ heterostructure with the polarization of In$_{2}$Se$_{3}$ pointing up and down, respectively. (f) denotes the top view of Fe$_{3}$GeTe$_{2}$/In$_{2}$Se$_{3}$ heterostructure, }
\end{figure*}

and time-reversal-odd part, denoting the intraband contribution to the torkances, which is an odd function of magnetization and is give by: 
\begin{equation}
\tau_{ij}^{\mathrm{odd}}
=
\frac{e\hbar}{\pi N}
\sum_{\mathbf{k},\,n,m}
\frac{
\Gamma^2
\operatorname{Re}
\left[
\langle n\mathbf{k}|\mathcal{T}_i|m\mathbf{k}\rangle
\langle m\mathbf{k}|v_j|n\mathbf{k}\rangle
\right]
}{
\left[
\left(E_{\textup{F}}-E_{\textup{n\textbf{k}}}\right)^2+\Gamma^2
\right]
\left[
\left(E_{\textup{F}}-E_{\textup{m\textbf{k}}}\right)^2+\Gamma^2
\right]
}.
\end{equation}

Here $|n\textbf{k}  \rangle$ and $E_{\textup{n\textbf{k}}}$ denotes the eigenstates and eigenvalues of Hamiltonian, $\textbf{k}$ and $n$ denotes the Bloch wave vector and band index, respectively. $E_{\textup{F}}$ denotes the Fermi level, $N$ is the total number used to sample k-space, and $\Gamma$ denotes the broadening parameter associated with the relaxation time $\tau$ by $\tau = \hbar/2\Gamma$. $\textbf{v}$ denotes the velocity operator, and $\mathcal{\boldsymbol{T}}$ denotes the torque operator, which is defined as $\mathcal{\boldsymbol{T}} = \frac{i}{\hbar} [H_{\textup{odd}},  \hat{\textbf{S}}]  $. $\hat{\textbf{S}}$ is the spin operator, while $H_{\textup{odd}}$ denotes the time-reversal-odd part of Wannier Hamiltonian\cite{brizolla2026anatomyspinorbittorquesmonolayer}. \textcolor{black}{For the calculation of angular dependence of torkance calculation, the exchange field related with $H_{\textup{odd}}$ is rotated manually from the ground state (+$z$ magnetization) to an arbitrary angle ($\theta$, $\phi$). Further details about the calculation procedure can be found in \cite{brizolla2026anatomyspinorbittorquesmonolayer}}. To further evaluate the torkance components contributed from a specific atom A, one just need to replace $\mathcal{\boldsymbol{T}}$ as $\mathcal{\boldsymbol{T}}^{\textup{A}}$, which is defined as $\mathcal{\boldsymbol{T}}^{\textup{A}} = \frac{1}{2}\ \{ \mathcal{T}, P^{\textup{A}} \}$\cite{Proj}. Here $P^{\textup{A}}$ is the projection operator onto the Wannier orbitals centered on atomic site A.

  \begin{figure*}[ht]
\includegraphics[scale = 0.4 ]{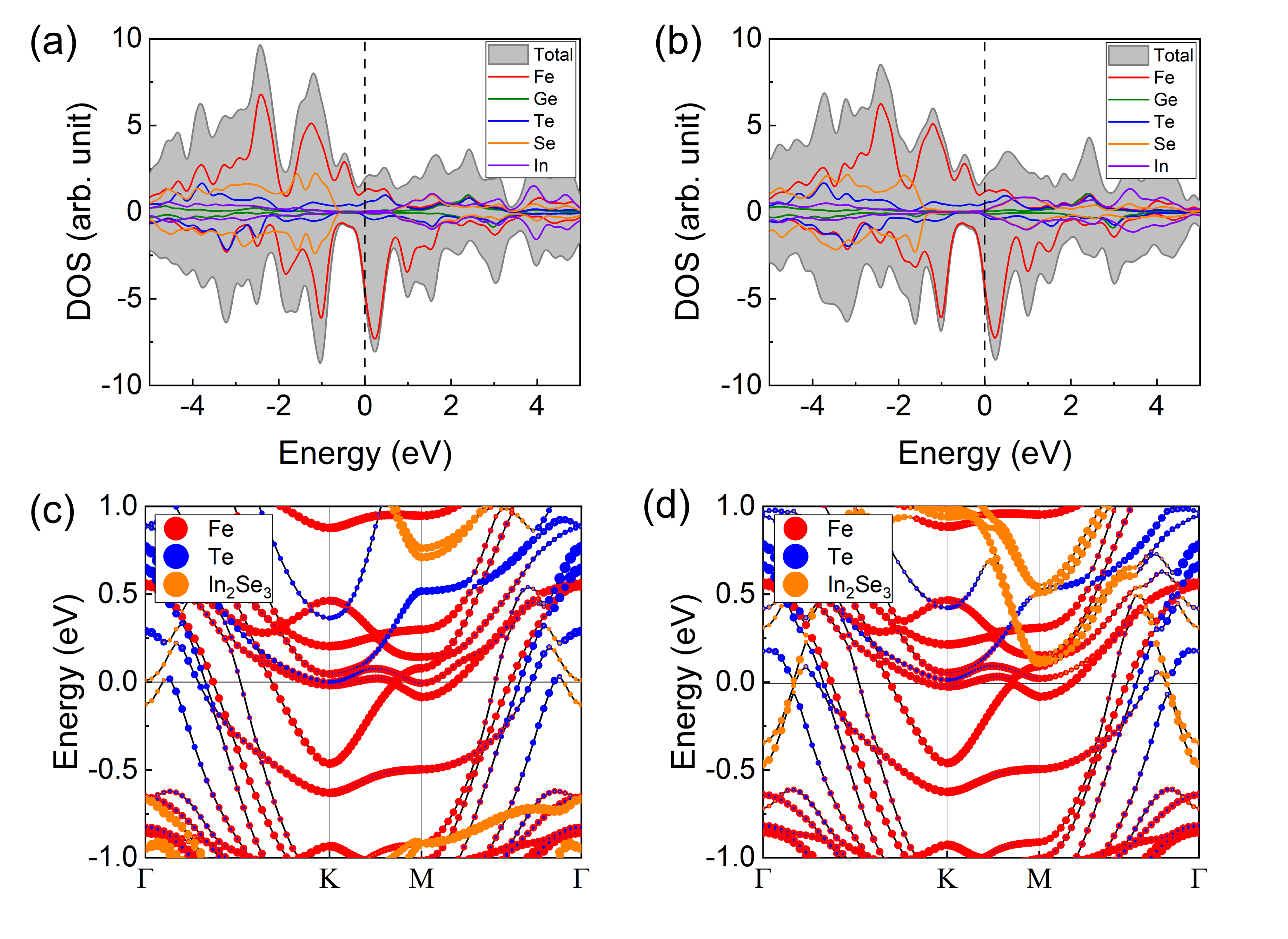}
\caption{\label{3} The calculated density of states for Fe$_{3}$GeTe$_{2}$/In$_{2}$Se$_{3}$ heterostructure with the polarization of In$_{2}$Se$_{3}$ (a) pointing upwards and (b) downwards, respectively. (c) and (d) shows the calculated projected band structure for heterostructure with the polarization of In$_{2}$Se$_{3}$ pointing upwards and downwards, respectively.}
\end{figure*}

In this work, we consider the case of electric field along +$x$ direction, i.e., $j$ = $x$. In this case, the calculated torkance components $\tau_{ix}$ is directly associated with the current-induced torque component along $i$ direction, $T_{i}$. Meanwhile, we define the magnitude of total torkance as 
\begin{equation}
    |\boldsymbol{\tau}| = \sqrt{ \tau_{xx} ^{2} +  \tau_{yx}^{2} + \tau_{zx}^{2} },
\end{equation}
which is associated with the magnitude of torque when the current is along +$x$ direction.

\textcolor{black}{In addition, the Edelstein effect is evaluated using the Linres\cite{linears} code within the Kubo-formula framework. Specifically, we consider the linear response of the spin polarization to an external electric field, expressed as $\delta \mathbf{S}=\chi\mathbf{E}$, where $\delta \mathbf{S}$ denotes the induced spin polarization, $\mathbf{E}$ is the electric field, and $\chi$ is the Edelstein response tensor. In this work we consider the time-reversal-even contribution to $\chi$, which can be written as \cite{Kubo1,pan2026tunableedelsteineffectintrinsic}}:

\begin{equation}
    \chi^{\textup{even}}_{ij} = -\frac{e\hbar}{\pi  N} \sum_{\mathbf{k},m,n} \frac{\Gamma^{2} \textup{Re}(\langle n\mathbf{k} |S_{i} | m\mathbf{k} \rangle \langle m\mathbf{k} | v_{j} | n\mathbf{k} \rangle  )  }{[(E_{F}-E_{n\mathbf{k}})^{2}+ \Gamma^{2}][(E_{F}-E_{m\mathbf{k}})^{2}+ \Gamma^{2}]}
\end{equation}

Here $\textbf{S}$ is the spin operator. For both calculation calculation of torkance and Edelstein effect, we set broaden parameter $\Gamma$ = 0.01 eV based on the previous work \cite{brizolla2026anatomyspinorbittorquesmonolayer}, which roughly corresponds to a typical sub-100 fs relaxation time. Meanwhile, $600\times600\times1$ k-mesh are adopted to arrive at the converged results.

\section{results and discussion}
\subsection{Basic electronic structure properties}
     \begin{figure*}[ht]
\includegraphics[scale = 0.38 ]{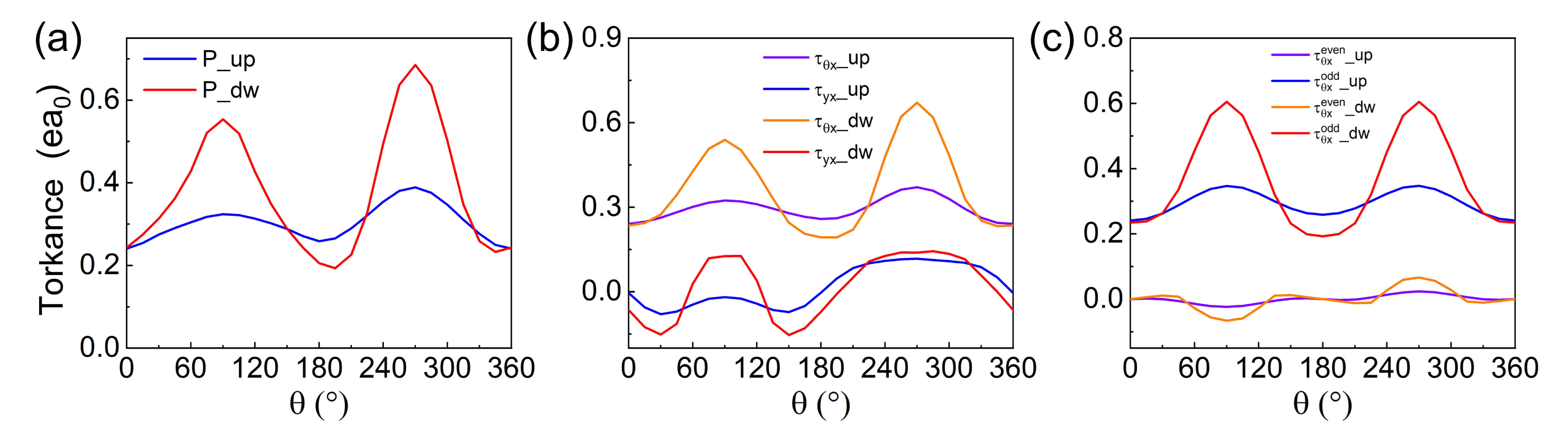}
\caption{\label{4} Angular-dependent torkance when magnetization rotating in $x-z$ plane. (a) The calculated magnitude of total torkance as function of magnetization angle $\theta$ for distinct polarization states in In$_{2}$Se$_{3}$ layer. (b) The calculated torkance components $\tau_{yx}$ and $\tau_{\theta x}$ as function of magnetization angle $\theta$ for distinct polarization states. (c) Calculated time-reversal-even and time-reversal-odd component of $\tau_{\theta x}$ as function of magnetization angle $\theta$ for distinct polarization states.}
\end{figure*}

The crystal structures of Fe$_{3}$GeTe$_{2}$ and In$_{2}$Se$_{3}$ are shown in Fig. \ref{2}. Monolayer Fe$_{3}$GeTe$_{2}$ is a ferromagnetic metal with out-of-plane magnetization, which has been successfully synthesized experimentally\cite{LDA2}. Its crystal structure consists of three Fe atomic layers sandwiched between two Te layers, with the Ge atoms and the central Fe (Fe2) atoms lying in the same atomic plane. This structure belongs to the non-centrosymmetric $D_{3h}$ point group, which allows the emergence of spin–orbit torque\cite{FGT}. The calculated magnetic moments of the Fe atoms are 1.82 $\mu_{B}$ for Fe1 and Fe3, and 1.05 $\mu_{B}$ for Fe2. The difference in magnetic moment between Fe1/Fe3 and Fe2 originates from their distinct local chemical environments and is in good agreement with previous reports\cite{LDA}. Furthermore, the magnetic anisotropy energy of monolayer Fe$_{3}$GeTe$_{2}$ is calculated to be $E_{x}$-$E_{z}$ = 3.57 meV per unit cell, indicating a strong out-of-plane magnetic anisotropy consistent with earlier studies\cite{LDA,xu2022assembling}. On the other hand, monolayer In$_{2}$Se$_{3}$ is a ferroelectric semiconductor\cite{In2Se3,In2Se32}, composed of five atomic sublayers stacked in the sequence Se–In–Se–In–Se and possessing $C_{3v}$ symmetry, which permits the existence of spontaneous out-of-plane electric polarization. When the electric polarization of monolayer In$_{2}$Se$_{3}$ is pointing upward along +$z$ direction (downward along -$z$ direction), the central Se layer shifts closer to the lower (upper) Se atomic layer, respectively.

The calculated in-plane lattice constants of monolayer Fe$_{3}$GeTe$_{2}$ and In$_{2}$Se$_{3}$ are 3.91 \AA and 3.98 \AA, respectively, resulting in a small lattice mismatch of approximately 1.7\%. Therefore, in constructing the Fe$_{3}$GeTe$_{2}$/In$_{2}$Se$_{3}$ heterobilayer, the in-plane lattice constant is fixed to 3.98 \AA, while all atoms are allowed to fully relax along the $z$ direction. To determine the lowest-energy stacking configuration, the total energies of different interlayer stackings were calculated as a function of relative interlayer sliding, as shown in the Appendix. The Fe$_{3}$GeTe$_{2}$/In$_{2}$Se$_{3}$ heterostructure with the lowest-energy stacking is presented in Figs. \ref{2}(d)-\ref{2}(f) and is used for subsequent calculations of the electronic structure and spin–orbit torque. It can be clearly seen that for both polarization states, the interfacial top Se atomic layer occupies the same in-plane positions as the Ge atoms in the Fe$_{3}$GeTe$_{2}$ layer. This heterostructure possesses $C_{3v}$ symmetry, which breaks the symmetry equivalence between the Fe1 and Fe3 atoms in the Fe$_{3}$GeTe$_{2}$ monolayer. This symmetry breaking is reflected in the calculated magnetic moments of Fe atoms for the Fe$_{3}$GeTe$_{2}$/In$_{2}$Se$_{3}$ heterostructure in the polarization-up state, yielding Fe1 = 1.88 $\mu_{B}$, Fe2 = 1.04 $\mu_{B}$, and Fe3 = 2.03 $\mu_{B}$, respectively. It is evident that the magnetic moment of Fe2 remains smaller than those of Fe1 and Fe3. However, in contrast to the monolayer Fe$_{3}$GeTe$_{2}$ case, the magnetic moments of Fe1 and Fe3 become inequivalent, which originates from their distinct local chemical environments induced by the heterostructure interface. Upon switching the polarization of In$_{2}$Se$_{3}$ from the upward to the downward state, the distribution of Fe magnetic moments remains nearly unchanged. The magnetic anisotropy energies of the Fe$_{3}$GeTe$_{2}$/In$_{2}$Se$_{3}$ heterostructure for the polarization-up and polarization-down states are calculated to be $E_{x} $ - $ E_{z}$ = 4.42 meV and 3.92 meV per unit cell, respectively. These results indicate that polarization switching does not alter the out-of-plane magnetization of the heterostructure system.

To further elucidate the electronic structure of the Fe$_{3}$GeTe$_{2}$/In$_{2}$Se$_{3}$ heterostructures, we calculate the density of states (DOS) for both the polarization-up and polarization-down states of In$_{2}$Se$_{3}$, as shown in Figs. \ref{3}(a) and \ref{3}(b). In both cases, pronounced exchange splitting is observed, accompanied by a sizable electronic density at the Fermi level, confirming the metallic nature of the heterostructure systems. The electronic states near the Fermi level are predominantly contributed by Fe and Te atoms, indicating strong hybridization between Fe and Te orbitals. For the polarization-up configuration, the contributions from Se and In atoms to the DOS near the Fermi level are negligible. In contrast, compared with the polarization-up case, the electronic states associated with Se and In atoms in the polarization-down heterostructure exhibit an overall downward energy shift. Such polarization-dependent modifications of the electronic states are also reflected in the band structures, shown in Figs. \ref{3}(c) and \ref{3}(d). The energy bands crossing the Fermi level are mainly derived from Fe and Te atoms, and these Fe- and Te-dominated bands remain robust against polarization switching. In addition, two energy bands originating from the In$_{2}$Se$_{3}$ layer appear near the $\Gamma$ point. Upon switching the polarization of In$_{2}$Se$_{3}$ from the upward to the downward state, these In$_{2}$Se$_{3}$-derived bands shift downward in energy, resulting in an expansion of the electronic pocket around the $\Gamma$ point. A similar downward shift of the In$_{2}$Se$_{3}$ conduction bands is also observed near the $\textup{M}$ point. Although the band structures presented here correspond to the magnetization oriented along the +$z$ direction, we note that the polarization-dependent features of band structures discussed above remain qualitatively unchanged when the magnetization is oriented along other directions, such as the +$x$ or +$y$, which is not shown here.

  \begin{figure*}[ht]
\includegraphics[scale = 0.38 ]{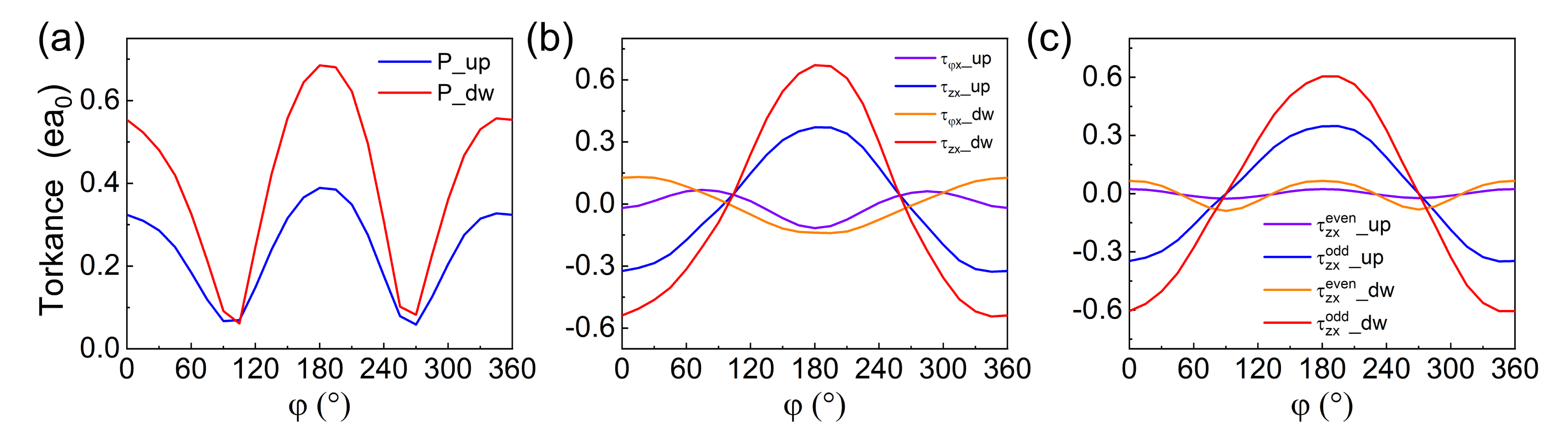}
\caption{\label{5}  Angular-dependent torkance when magnetization rotating in $x-y$ plane. (a) The calculated magnitude of total torkance as function of magnetization angle $\varphi$ for distinct polarization states in In$_{2}$Se$_{3}$ layer. (b) The calculated torkance components $\tau_{zx}$ and $\tau_{\varphi x}$ as function of magnetization angle $\varphi$ for distinct polarization states. (c) Calculated time-reversal-even and time-reversal-odd component of $\tau_{z x}$ as function of magnetization angle $\theta$ for distinct polarization states.}
\end{figure*}

\subsection{Polarization dependent spin-orbit torque}

The polarization-dependent modification of the electronic structure is expected to alter the SOT in the heterostructure, which also depends sensitively on the magnetization direction. To this end, we first evaluate the angular dependence of the torkance by considering electric field along +$x$ direction and magnetization rotation between the out-of-plane and in-plane directions, taking the case where the magnetization lies in the $x-z$ plane as a representative example. In this configuration, the torkance can be expressed as a function of the polar angle $\theta$, defined as the angle between the magnetization direction and the +$z$ axis. As shown in Fig. \ref{4}(a), both the polarization-up and polarization-down heterostructures exhibit a clear variation in the magnitude of total torkance defined in Eq.(4) as the magnetization direction changes.

\textcolor{black}{When magnetization is colinear with $z$ axis ($\theta = 0^\circ$ and $180^\circ$), the torkance, which reaches its minima value, is finite for both polarization states. This behavior contrasts with that of monolayer Fe$_{3}$GeTe$_{2}$, where the torkance vanishes for +$z$ or -$z$ magnetization due to the constraint imposed by $D_{3h}$ symmetry\cite{FGT}. As the magnetization direction continuously rotates from the out-of-plane to the in-plane orientation, the total torkance in both heterostructures exhibits an overall increasing trend and reaches its maximum when the magnetization is collinear with the $x$ axis ($\theta = 90^\circ$ and $270^\circ$). Despite this overall similarity, pronounced quantitative differences emerge in the angular dependence of the torkance between the two polarization states. Although the total torkances of the two polarization states are comparable when the magnetization is oriented along the $+z$ or $-z$ direction, the torkance in the polarization-down state increases more rapidly than that in the polarization-up state as the magnetization rotates from the out-of-plane to the in-plane orientation. Consequently, the difference in total torkance between the two polarization states becomes largest when the magnetization lies in the $x-y$ plane. Quantitatively, at $\theta = 90^\circ$ ($270^\circ$), switching the polarization from up to down increases the total torkance by 0.22 (0.30) $e a_{0}$, so that the total torkance reaches approximately 170\% (180\%) of its original value.}

To gain deeper insight into the origin of the polarization-dependent variation of the torkance, it is necessary to analyze how individual torkance components evolve under different polarization states. For this purpose, we consider two torkance components, $\tau_{yx}$ and $\tau_{\theta x}$. Here $\tau_{\theta x}$ is defined as $\tau_{\theta x} = \tau_{xx}\textup{cos}\theta - \tau_{zx}\textup{sin}\theta$. When current is flowing along +$x$ direction, these two torkance components are associated with torque components along $y$ direction and $\theta$-denoted tangential direction denoted as $\hat{\textbf{e}}_{\theta} = (\textup{cos}\theta, 0 ,\textup{sin}\theta)^{\textup{T}}$, which play a dominant role in governing the magnetization dynamics when the magnetization lies in the $x–z$ plane. The angular dependence of the corresponding torkance components, $\tau_{yx}$ and $\tau_{\theta x}$, under different polarization states is shown in Fig. \ref{4}(b). As can be seen, for both polarization-up and polarization-down states, $\tau_{yx}$ and $\tau_{\theta x}$ reach their local maximum values at $\theta = 90^\circ$ and $270^\circ$, which is consistent with the observation that the total torkance of the heterostructure attains its maximum at this angle. Moreover, the magnitude of $\tau_{\theta x}$ is always larger than that of $\tau_{yx}$, indicating that the total torkance is predominantly governed by the $\tau_{\theta x}$ component. Interestingly, upon switching the polarization from the upward to the downward state, the angular dependence of $\tau_{\theta x}$ exhibits a pronounced variation, whereas $\tau_{yx}$ remains relatively unchanged. The most significant change in $\tau_{\theta x}$ occurs at $\theta = 90^\circ$ and $270^\circ$, where its magnitude increases by approximately $0.2$ and $0.3 ea_{0}$, respectively. Upon polarization switching, $\tau_{\theta x}$ reaches about 160\% and 180\% of its original value at $\theta = 90^\circ$ and $270^\circ$, respectively. Based on these observations, we conclude that when the magnetization lies in the $x–z$ plane with current along +$x$, the polarization-dependent modulation of the torkance in the Fe$_{3}$GeTe$_{2}$/In$_{2}$Se$_{3}$ heterostructure is primarily dominated by the $\tau_{\theta x}$.

To further elucidate the polarization dependence of $\tau_{\theta x}$, which dominate the total torkance, we decompose it into a time-reversal-even component $\tau^{\textup{even}}_{\theta x}$ and a time-reversal-odd component $\tau^{\textup{odd}}_{\theta x}$. As introduced in the Methods section, this decomposition enables one to distinguish between the interband and intraband contributions to the torkance. As shown in Fig. \ref{4}(c), for both polarization states, $\tau^{\textup{odd}}_{\theta x}$ is consistently more than one order of magnitude larger than $\tau^{\textup{even}}_{\theta x}$, indicating that $\tau_{\theta x}$ is predominantly governed by the time-reversal-odd contribution. Moreover, upon switching the polarization from the upward to the downward state, the time-reversal-odd torkance $\tau^{\textup{odd}}_{\theta x}$ exhibits a much more pronounced variation compared with the time-reversal-even component. The maximum change occurs at $\theta = 90^\circ$ and $270^\circ$, where $\tau^{\textup{odd}}_{\theta x}$ increases by approximately $0.25 ea_{0}$, reaching about 170\% of its original value upon polarization switching.

In addition to the case where the magnetization rotates from the out-of-plane to the in-plane direction, as discussed above, we also investigate the angular-dependent torkance for magnetization rotation within the $x–y$ plane. The results of angular-dependent total torkance variation are shown in Fig. \ref{5}(a). Here, the angle $\varphi$ is defined as the azimuthal angle between the magnetization direction and the $+x$ axis within the $x–y$ plane, which can be seen in Fig. \ref{1}(a). As $\varphi$ varies from $0^\circ$ to $180^\circ$, the total torkance for both polarization states initially decreases and then increases, with the maximum and minimum values occurring at $\varphi = 180^\circ$ and $\varphi = 90^\circ$, respectively. Although the two polarization states exhibit a similar overall evolution trend, their torkance magnitudes differ noticeably. At $\varphi = 0^\circ$ and $ 180^\circ$, which corresponds to magnetization along $+x$ and $-x$, the difference in the total torkance between two polarization configuration reaches 0.23 and 0.3 $ea_{0}$, respectively.

To further clarify the underlying mechanism, we further consider two torkance components:  $\tau_{z x}$ and $\tau_{\varphi x}$. Here $\tau_{\varphi x}$ is defined as $\tau_{\varphi x} = -\tau_{x x} \textup{sin} \varphi +  \tau_{y x} \textup{cos} \varphi  $. For current along +$x$ direction, these two torkance components are associated with torque components along $z$ direction and $\varphi$-denoted tangential direction denoted as $\hat{\textbf{e}}_{\varphi} = (-\textup{sin}\varphi, \textup{cos}\varphi, 0)^{\textup{T}}$, which play a dominant role in governing the magnetization dynamics when the magnetization lies in the $x–y$ plane.
The calculated angular dependence of these torkance components is presented in Fig.  \ref{5}(b). It can be seen that for both polarization-up and polarization-down configurations, the torkance is predominantly governed by the $\tau_{z x}$ component. Such $\tau_{z x}$ component exhibits a pronounced variation of approximately $0.2  ea_{0}$ upon polarization switching when $\varphi$ is close to $0^\circ$ or $180^\circ$. In contrast, the $\tau_{\varphi x}$ component is much smaller in magnitude compared with $\tau_{z x}$, and its overall variation against polarization switching is also weaker than that of $\tau_{z x}$. Therefore, it is fair to say that when the magnetization lies in the $x–y$ plane, the polarization-dependent variation of the total torkance in the Fe$_{3}$GeTe$_{2}$/In$_{2}$Se$_{3}$ heterostructure is dominated by the $\tau_{z x}$ component.

     \begin{figure*}[ht]
\includegraphics[scale = 0.3 ]{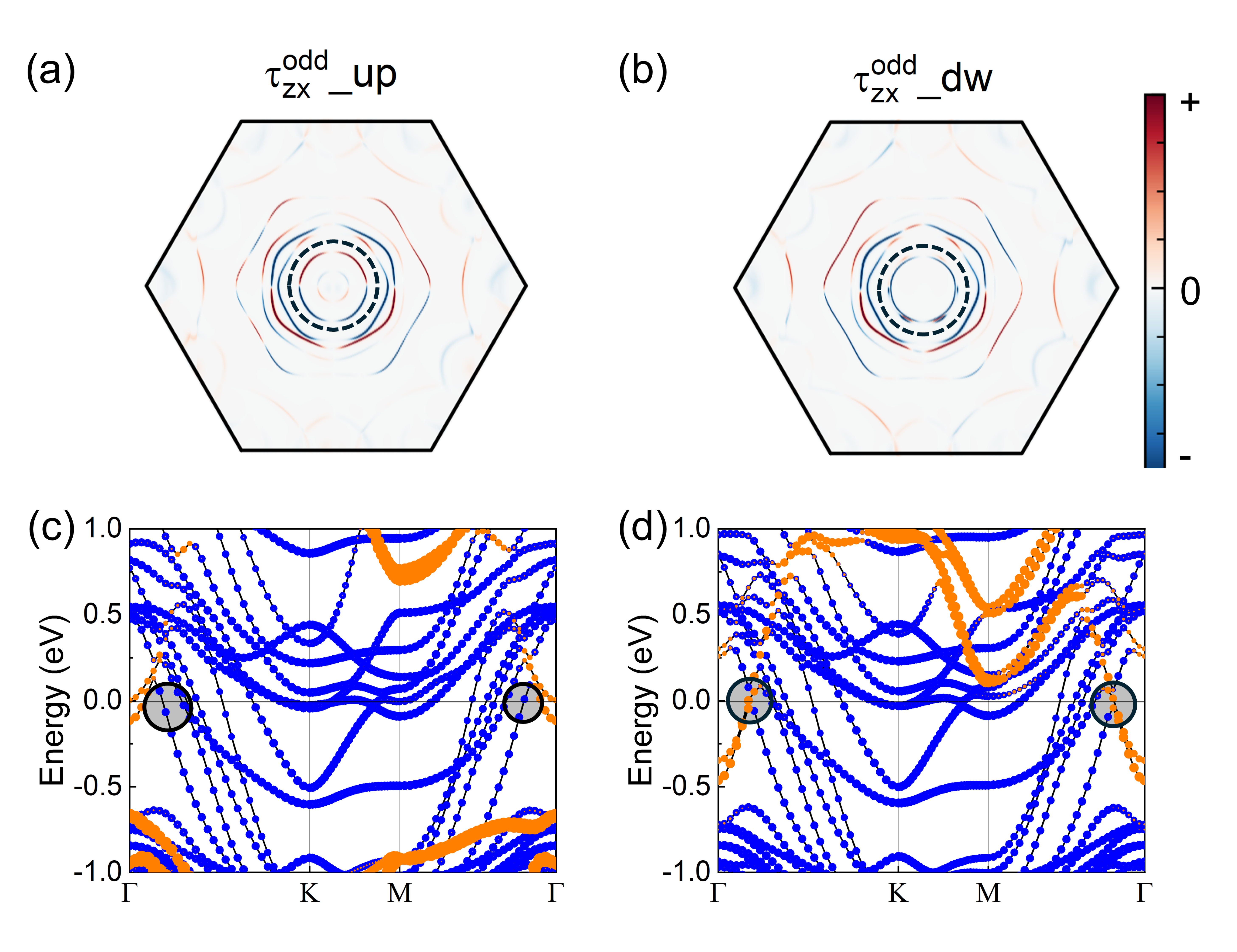}
\caption{\label{6} The k-resolved polarization-dependent $\tau^{\textup{odd}}_{z x}$ in the first Brillouin zone for (a) polarization-up and (b) polarization-down, respectively. $600\times600$ kmesh and $\Gamma$ = 0.01 eV is adopted. Red and blue denotes the positive and negative contribution to the torkance ($\pm$ 60.2 for (a) and $\pm$ 66.3 for (b)). The area of interest, which includes the arc-like contribution of torkance, is circled.  (c) and (d) shows the projected band structure for (c) polarization-up and (d) polarization-down, respectively. Blue denotes the electronic states from Fe$_{3}$GeTe$_{2}$, while orange denotes that from In$_{2}$Se$_{3}$. The gray dashes denote the electronic states on the Fermi surface that contribute to the arc-like torkance distributions circled in panels (a) and (b).
}
\end{figure*}

    \begin{figure*}[ht]
\includegraphics[scale = 0.38 ]{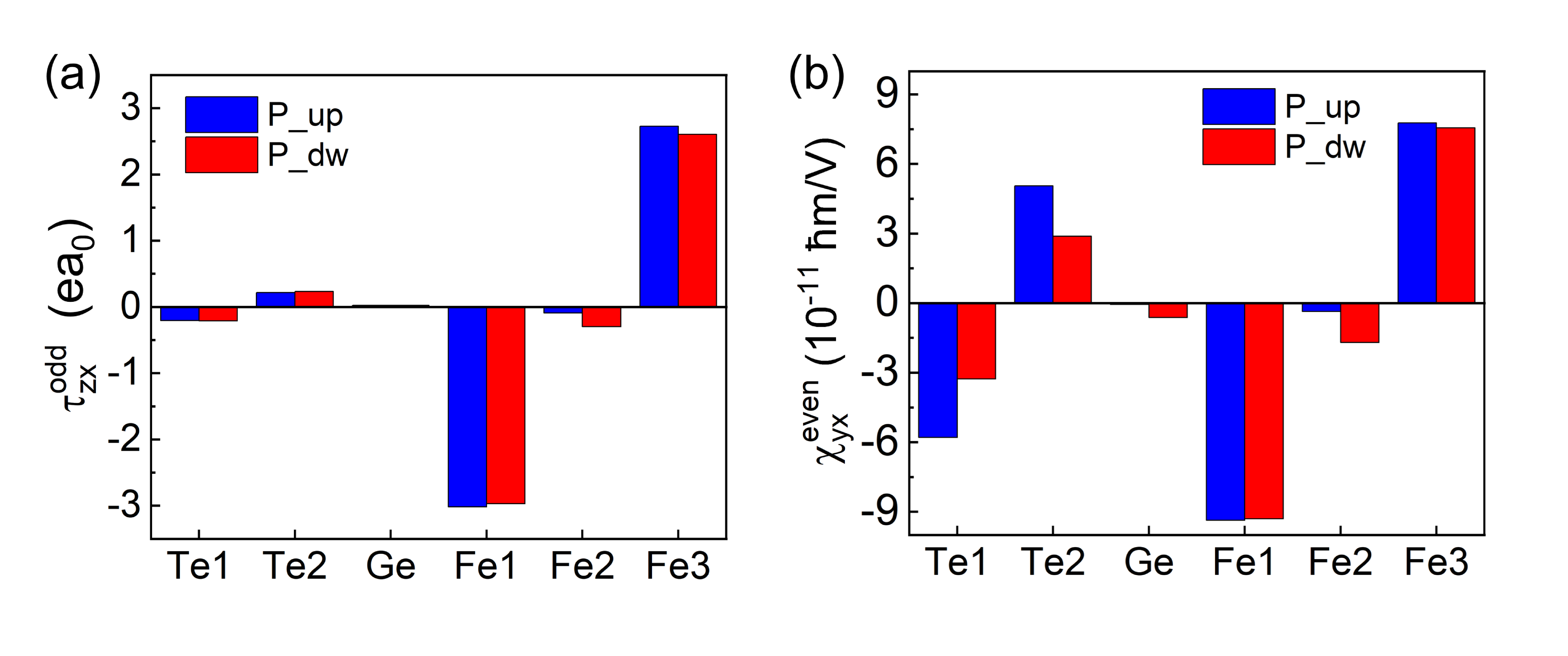}
\caption{\label{9} The atom-resolved (a) $\tau^{\textup{odd}}_{z x}$ and (b) $\chi^{\textup{even}}_{z x}$ for Fe$_{3}$GeTe$_{2}$/In$_{2}$Se$_{3}$ heterostructure with distinct polarizations. The label of atoms can be seen in Fig. \ref{2}(a). Here we only show the contributions from atoms on the Fe$_{3}$GeTe$_{2}$ layer, while the torkance on the atoms from In$_{2}$Se$_{3}$ layer are negligible.
}
\end{figure*}

To gain deeper insight into the polarization- and angle-dependent evolution of $\tau_{z x}$, which plays the dominant role in the total torkance variation against polarization switching, we decompose it into time-reversal-even and time-reversal-odd components (denoted as $\tau^{\textup{even}}_{z x}$ and $\tau^{\textup{odd}}_{z x}$ respectively), and plot them as functions of the polarization state and the azimuthal angle $\varphi$ in Fig.  \ref{5}(c). It is evident that the time-reversal-odd component $\tau^{\textup{odd}}_{z x}$ is significantly larger in magnitude than the time-reversal-even component $\tau^{\textup{even}}_{z x}$, and thus dominates $\tau_{z x}$ in both polarization states. Moreover, upon polarization switching, $\tau^{\textup{odd}}_{z x}$ component exhibits a pronounced variation, whereas the $\tau^{\textup{even}}_{z x}$ component remains close to zero and undergoes only a minor change. In particular, when $\varphi=0^\circ$ or ${180}^\circ$, $\tau^{\textup{odd}}_{z x}$ shows its maximum polarization-induced variation of approximately $0.25 ea_{0}$.

\subsection{K-resolved torkances with distinct polarizations}

The above analysis reveals that when magnetization is pointing along +$x$ direction, the dominant time-reversal-odd torkance component, $\tau^{\textup{odd}}_{zx}$ exhibits a pronounced variation upon polarization reversal. To gain deeper insight into the microscopic origin of this behavior, we further decompose $\tau^{\textup{odd}}_{zx}$ into its k-resolved contributions for the polarization-up and polarization-down states.
As shown in Figs.\ref{6}(a) and \ref{6}(b), arc-like torkance distributions emerge in momentum space for both polarization states. Because the time-reversal-odd torkance is dominated by electronic states at the Fermi surface, the shapes of these arc-like features closely follow the distribution of the Fermi-surface states. The torkance contributions are particularly prominent near the $\Gamma$ point, where they exhibit an alternating pattern of positive and negative values.
Notably, polarization reversal substantially modifies the torkance distribution in the vicinity of the $\Gamma$ point. In the polarization-up state, the second ring counted outward from the $\Gamma$ point displays alternating positive and negative contributions, as highlighted by the dashed circle in Fig. \ref{6}(a). By contrast, in the polarization-down state, the ring at a comparable radius is dominated entirely by negative contributions, as highlighted in Fig. \ref{6}(b). Consequently, switching the polarization from the up state to the down state introduces additional negative contributions to the k-resolved torkance. This redistribution in momentum space is consistent with the observed variation of $\tau^{\textup{odd}}_{zx}$ for magnetization oriented along the $+x$ direction.

Microscopically, the redistribution of the torkance in momentum space is closely associated with polarization-induced modifications of the electronic structure at the Fermi surface, as shown in Figs. \ref{6}(c) and \ref{6}(d). In the polarization-up configuration, an Fe$_{3}$GeTe$_{2}$-derived electronic state near the $\Gamma$ point forms a hole pocket at the Fermi surface. The electronic states forming this hole pocket give rise to the arc-like $\tau^{\textup{odd}}_{zx}$ contributions with alternating positive and negative signs, as highlighted by the circle in Fig. \ref{6}(a).
Upon switching the polarization from the up to the down configuration, the In$_{2}$Se$_{3}$-derived bands shift downward in energy. Near the $\Gamma$ point, the In$_{2}$Se$_{3}$-derived and Fe$_{3}$GeTe$_{2}$-derived states consequently hybridize and form an anticrossing at the Fermi surface. This hybridization transforms the original Fe$_{3}$GeTe$_{2}$-derived hole pocket into an In$_{2}$Se$_{3}$-derived electron pocket, whose constituent electronic states give rise to the all-negative arc-like $\tau^{\textup{odd}}_{zx}$ contributions highlighted in Fig. \ref{6}(b). Therefore, the polarization-induced reconstruction of the Fermi-surface electronic structure near the $\Gamma$ point is consistent with the corresponding change in the arc-like distribution of $\tau^{\textup{odd}}_{zx}$, ultimately giving rise to the variation in the total $\tau_{zx}$.

\subsection{Atom-resolved torkances and Edelstein effect with distinct polarizations}
Polarization reversal not only modifies the k-space distribution of $\tau^{\textup{odd}}_{zx}$, but also alters its atom-resolved contributions. To gain deeper insight into this effect, we further decompose $\tau^{\textup{odd}}_{zx}$ into atom-resolved contributions for the polarization-up and polarization-down states. As shown in Fig. \ref{9}(a), in both polarization configurations, Fe1 and Fe3 provide the dominant contributions to the torkance. However, their contributions have opposite signs and therefore largely cancel each other, whereas the contribution from Fe2 is smaller but remains non-negligible. \textcolor{black}{We note that atom-resolved torkance has also been reported for pristine Fe$_{3}$GeTe$_{2}$. Similar to the Fe$_{3}$GeTe$_{2}$/In$_{2}$Se$_{3}$ heterostructure, the out-of-plane time-reversal-odd torkance contributions from the Fe1 and Fe3 atoms have opposite signs and therefore cancel when the magnetization is oriented along the $+x$ direction \cite{brizolla2026anatomyspinorbittorquesmonolayer}. However, their magnitudes in pristine Fe$_{3}$GeTe$_{2}$ are approximately one order of magnitude smaller than those in the heterostructure. Moreover, the corresponding torkance contribution from the Fe2 atoms remains zero in the pristine Fe$_{3}$GeTe$_{2}$ monolayer\cite{brizolla2026anatomyspinorbittorquesmonolayer}, in marked contrast to the nontrivial Fe2 contribution found in Fe$_{3}$GeTe$_{2}$/In$_{2}$Se$_{3}$. These observations demonstrate that the proximity coupling between Fe$_{3}$GeTe$_{2}$ and In$_{2}$Se$_{3}$ substantially modifies the local torkance on the magnetic Fe atoms.}
Besides Fe atoms, The Te atoms also make finite contributions to the torkance, possibly owing to the combined effects of their strong spin–orbit coupling and the magnetic moments induced on Te through hybridization with the Fe states. Nevertheless, the contributions from Te1 and Te2 have opposite signs and nearly cancel, leading to an almost negligible net contribution from Te. Upon switching the polarization from the upward to the downward state, the contributions from Te1, Te2, Ge, Fe1, and Fe3 change only slightly. In sharp contrast, the Fe2 contribution is substantially enhanced in magnitude, increasing from -0.09 to -0.30 $ea_{0}$, corresponding to a 233\% increase in magnitude. These results indicate that the pronounced polarization dependence of the Fe2 contribution plays a key role in modulating the $\tau^{\textup{odd}}_{zx}$.

Besides the atom-resolved SOT, polarization reversal also modifies the distribution of the current-induced spin polarization, which is generally closely related to the generation of SOT and are usually used for a qualitative estimation for. To gain further insights, we therefore calculate the atom-resolved Edelstein coefficient for both polarization states. Here, the Edelstein response refers to the time-reversal-even Fermi-surface contribution to the current-induced $y$-component of the spin polarization under an electric field applied along the $+x$ direction, denoted by $\chi_{yx}^{\textup{even}}$. Through its cross product with the magnetization oriented along the $+x$ direction, this transverse spin response generates a $z$-component torque that corresponds with the time-reversal-odd torkance component $\tau^{\textup{odd}}_{zx}$.
As shown in Fig. \ref{9}(b), the Fe-resolved Edelstein coefficient qualitatively follows the distribution of the atom-resolved torkance. Fe1 and Fe3 provide the dominant contributions with opposite signs, whereas the Fe2 contribution is substantially smaller. Upon polarization reversal, the Edelstein responses of Fe1 and Fe3 vary only weakly, while that of Fe2 changes significantly. These features are qualitatively consistent with the polarization dependence of the Fe-resolved torkance. By contrast, the Edelstein coefficients on the Te sites do not exhibit the same relative changes as their torkance contributions.

We now work to understand the mismatch between Edelstein coefficient and torkance. Microscopically, the atom-resolved Edelstein coefficient is proportional to the current-induced effective magnetic field \cite{RevModPhys.91.035004}, which can be obtained by integrating the electric-field-induced spin response over the atomic region,

\begin{equation}
    \delta M^{(i)}=\int_{i}\delta M(\mathbf{r})\,\mathrm{d}\mathbf{r}
\end{equation}

whereas the atom-resolved torkance is proportional to the current-induced torque, which is determined by the local exchange-field-weighted spin response\cite{Kubo1},

\begin{equation}
T^{i}=\int_{i}\mathbf{\Omega}_{\textup{xc}}(\mathbf{r}) \times \delta M(\mathbf{r})\,\mathrm{d}\mathbf{r}
\end{equation}

If the exchange field $\mathbf{\Omega}_{\textup{xc}}(\mathbf{r})$ were spatially uniform within each atomic region, one could approximately get: 

\begin{equation}
   T^{i}\approx\mathbf{\Omega}_i\times \delta M ^{\left(i\right)} 
\end{equation}

where $\mathbf{\Omega}_i = \int_{i} \mathbf{\Omega}_{\textup{xc}}(\mathbf{r})$ is the integrated exchange field on each atom. 
In this case, if the local exchange field remained unchanged upon polarization reversal, the atom-resolved Edelstein coefficient and torkance would be expected to exhibit the same relative variation under polarization switching. These assumptions, however, are not generally valid for an itinerant magnet such as Fe$_{3}$GeTe$_{2}$, in which the exchange field is largely non-uniform distributed on each atom. Moreover, polarization reversal modifies the electronic structure and can alter both the magnitude and the spatial distribution of the exchange field and nonequilibrium spin density within an atomic region. Consequently, the polarization dependence of the atom-resolved Edelstein response need not coincide quantitatively with that of the atom-resolved torkance. These results demonstrate that atom-resolved torkance cannot, in general, be inferred directly from the atom-resolved Edelstein coefficient. An explicit evaluation of the microscopic torkance is therefore required to precisely determine the torque contribution from each atomic site.

   \begin{figure*}[ht]
\includegraphics[scale = 0.33 ]{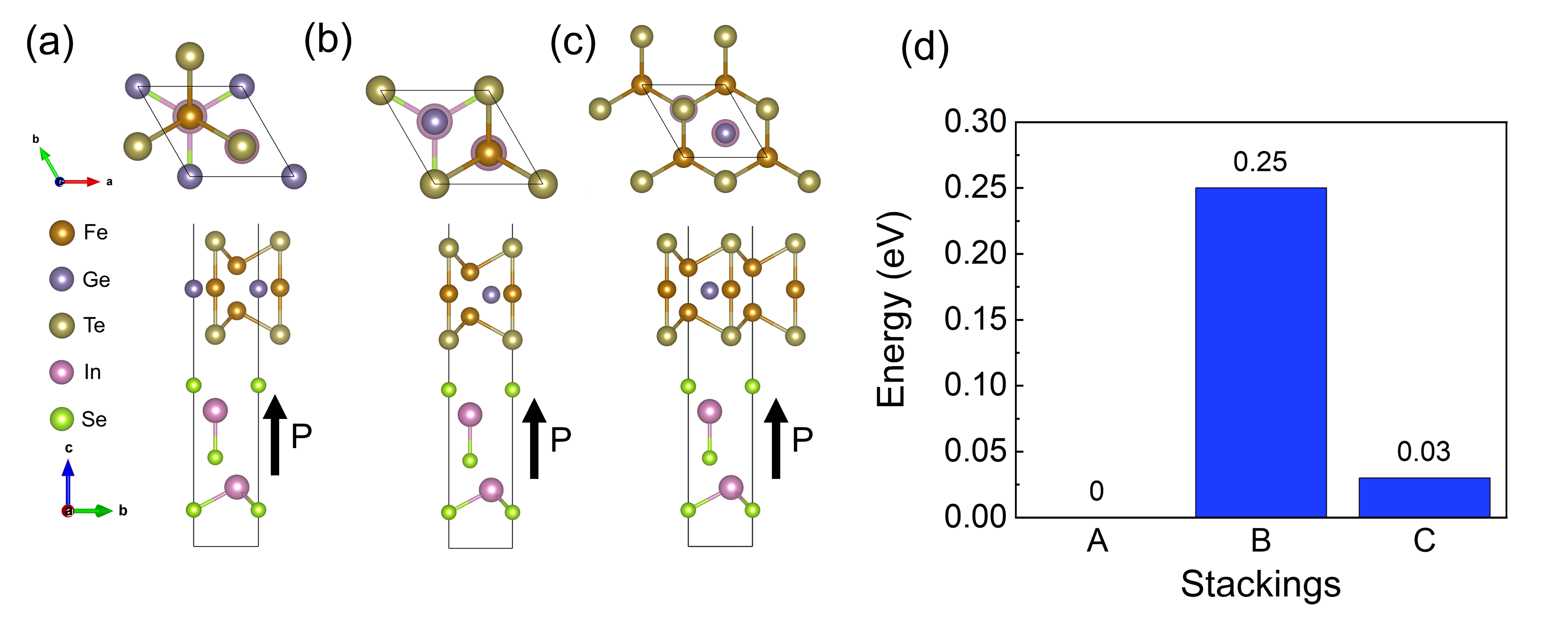}
\caption{\label{7} (a)-(c) denotes three atomic configurations, A, B, and C, with polarization-up, which are used for evaluating the energy. The calculated energies are shown in (d) with respect to the lowest energy configuration A. }
\end{figure*}

\section{
Summary}

In summary, based on first-principles calculations, we have systematically investigated ferroelectrically controllable SOT in Fe$_{3}$GeTe$_{2}$/In$_{2}$Se$_{3}$ heterostructures. We find that reversing the ferroelectric polarization of In$_{2}$Se$_{3}$ leads to a pronounced modulation of the angular dependence of torkance, with the torkance reaching up to more than 150\% of its original value when the magnetization is oriented along the +$x$ direction. Further analysis reveals that, for such +$x$ magnetization configuration, the polarization-dependent variation of the total torkance is dominated by the $z$-component of time-reversal-odd torkance, which is mainly attributed to the redistribution of electronic structure on the Fermi surface. In addition, atomic resolved analysis of torkance shows the middle layer of Fe atoms in Fe$_{3}$GeTe$_{2}$ layer dominately contributes to the variation of torkance upon polarization switching. Our findings demonstrate that ferroelectric control provides an effective route for nonvolatile manipulation of SOT in van der Waals multiferroic heterostructures. We believe that this work not only offers new insights into the functional potential of vdW multiferroic systems, but also paves the way toward the realization of nonvolatile and energy-efficient SOT-based spintronic devices.

\section{Acknowledgements}
This project was supported by the European Union Graphene Flagship project 2DSPIN-TECH (grant agreement No. 101135853) and SFB 1277 (Project-ID 314695032).

\section{Appendix A: Energies of distinct stacking orders for the heterostructures.}

   \begin{figure*}[ht]
\includegraphics[scale = 0.33 ]{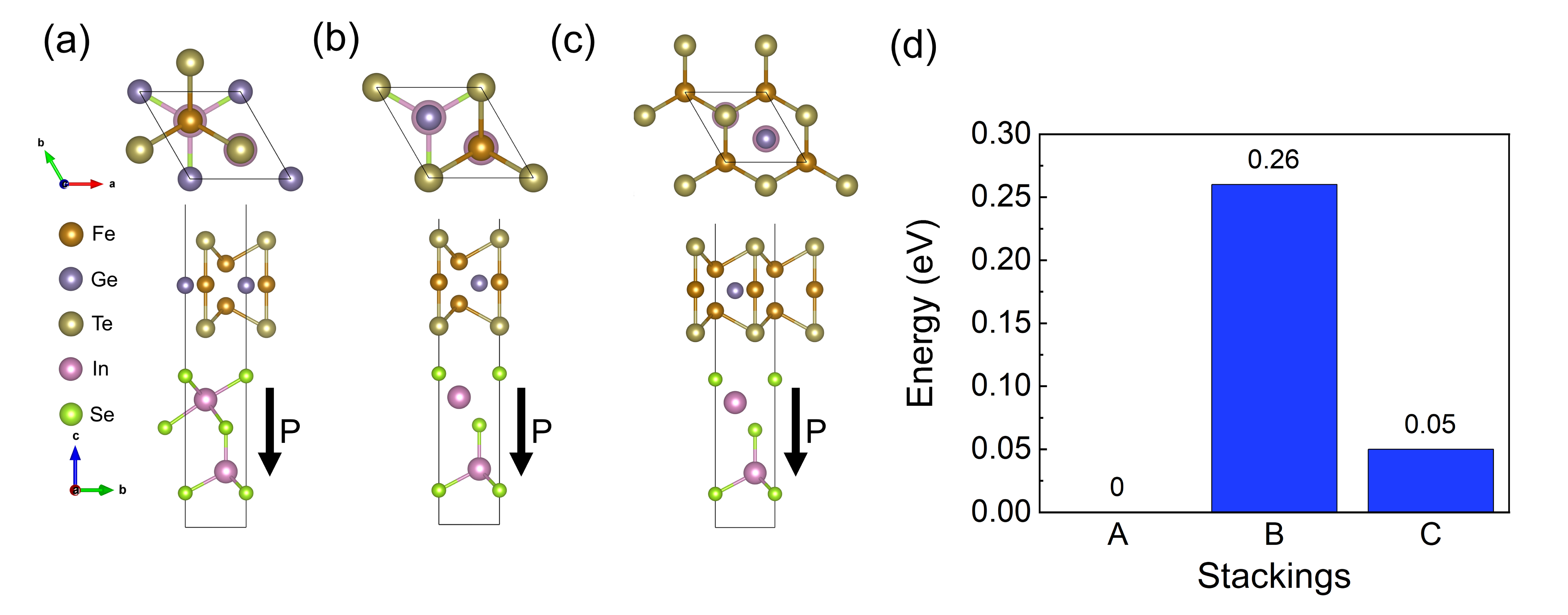}
\caption{\label{8} (a)-(c) denotes three atomic configurations, A, B, and C, with polarization-down, which are used for evaluating the energy. The calculated energies are shown in (d) with respect to the lowest energy configuration A. }
\end{figure*}
  
To determine the lowest-energy stacking configuration of the Fe$_{3}$GeTe$_{2}$/In$_{2}$Se$_{3}$ heterostructure, we start from an initial stacking arrangement in which the Ge atoms are positioned directly above the nearest In atoms, denoted as configuration $\textup{A}$. Subsequently, the In$_{2}$Se$_{3}$ layer is kept fixed while the Fe$_{3}$GeTe$_{2}$ layer is laterally shifted along the $\langle 1\overline{1}0 \rangle$ direction by one-third and two-thirds of the unit-cell lattice vector, leading to configurations $\textup{B}$ and $\textup{C}$, respectively. It is worth noting that all three configurations—$\textup{A}$, $\textup{B}$, and $\textup{C}$—preserve $C_{3v}$ symmetry, and are therefore expected to be potential candidates for low-energy stacking arrangements. By calculating the total energies of these configurations, we find that configuration $\textup{A}$ has the lowest energy for both polarization-up and polarization-down states of In$_{2}$Se$_{3}$, which is shown in Figs. \ref{8} and \ref{9}. Consequently, configuration $\textup{A}$ is adopted in the main text for all subsequent calculations and analyses.

\section{Appendix B: Angular dependence of torkance}

    \begin{figure*}[ht]
\includegraphics[scale = 0.33 ]{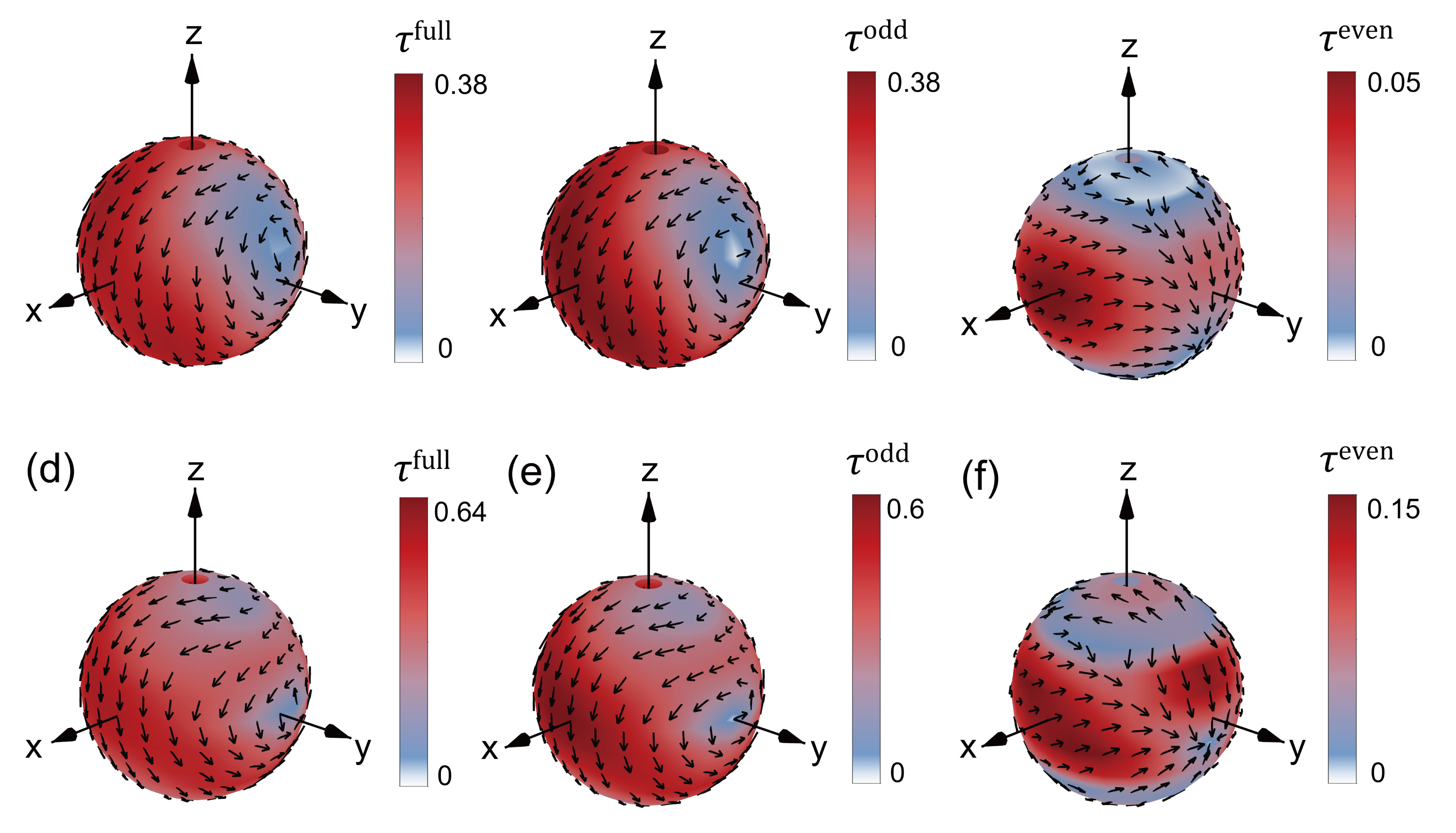}
\caption{\label{10} Angular dependence of the torkance on the magnetization direction ($\theta, \phi$) when the electric field is along +$x$. For polarization-up configuration, (a)-(c) shows the full torkance, time-reversal-odd torkance, and time-reversal-even torkance, respectively.  For polarization-down configuration, (d)-(f) shows the full torkance, time-reversal-odd torkance, and time-reversal-even torkance, respectively. The arrow on the sphere indicates the direction, while the color denotes the magnitude. The unit of the torkance is $ea_{0}$. 
}
\end{figure*}

\textcolor{black}{To obtain a global qualitative view of the angular-dependent torkance for distinct polarization configurations, we plot its distribution over the full sphere of magnetization orientations for both polarization states, as shown in Fig. \ref{10}. For both polarization configurations, the total torkance reaches its maximum for magnetization along the $+x$ direction and exhibits a local minimum when the magnetization lies in the $yz$ plane. Meanwhile, the torkance remains finite at both the north and south poles of the sphere. This behavior contrasts with that of monolayer Fe$_{3}$GeTe$_{2}$, where the $D_{3h}$ symmetry enforces a vanishing total torkance at both poles. Moreover, the time-reversal-odd torkance is approximately one order of magnitude larger than its time-reversal-even counterpart and therefore dominates the total response. The angular dependence of the time-reversal-odd torkance closely resembles that of the conventional field-like torque, $\mathbf{y}\times\mathbf{m}$, suggesting that the field-like contribution may dominate the time-reversal-odd component. In contrast, the angular dependence of the time-reversal-even torkance deviates markedly from the conventional damping-like form, $\mathbf{m}\times(\mathbf{y}\times\mathbf{m})$, indicating the presence of higher-order angular contributions beyond damping-like formula, which is not the focus of the present work. When the polarization is switched from up to down, the angular dependence of the torkance remains qualitatively similar, whereas the magnitude of the dominant time-reversal-odd component is strongly enhanced. }

%

\end{document}